\documentclass[a4paper,11pt]{article}

\usepackage{jinstpub} 

\title{\boldmath Continuous pulse advances in the negative ion source NIO1}

\author[a]{M. Barbisan,\note{Corresponding author.}}
\author[a,b]{R. Agnello,}
\author[c]{M. Cavenago,}
\author[a]{R. S. Delogu,}
\author[a]{A. Pimazzoni,}
\author[d]{L. Balconi,}
\author[a]{P. Barbato,}
\author[a]{L. Baseggio,}
\author[e]{A. Castagni,}
\author[a,b]{B. Pouradier Duteil,}
\author[a]{L. Franchin,}
\author[a]{B. Laterza,}
\author[a]{F. Molon,}
\author[a]{M. Maniero,}
\author[a]{L. Migliorato,}
\author[a]{R. Milazzo, }
\author[a]{G. Passalacqua,}
\author[a]{C. Poggi,}
\author[a]{D. Ravarotto,}
\author[a]{R. Rizzieri,}
\author[a]{L. Romanato,}
\author[a]{F. Rossetto,}
\author[a]{L Trevisan,}
\author[c,a]{M. Ugoletti,}
\author[a]{B. Zaniol}
\author[a]{and S. Zucchetti}

\affiliation[a]{Consorzio RFX (CNR, ENEA, INFN, Università di Padova, Acciaierie Venete SpA), C.so Stati Uniti 4, 35127 Padova, Italy}
\affiliation[b]{EPFL, Swiss Plasma Center (SPC), CH-1015 Lausanne, Switzerland}
\affiliation[c]{INFN-LNL, v.le dell’Università n. 2, I-35020, Legnaro (PD) Italy}
\affiliation[d]{Università degli Studi di Milano, Dipartimento di Fisica, via Celoria 16, 20122 Milano, Italy }
\affiliation[e]{Università di Modena e Reggio Emilia, Dipartimento di Scienze Fisiche, Informatiche e Matematiche, Via Campi 213/A - 41125 Modena, Italy }

\emailAdd{marco.barbisan@igi.cnr.it}

\abstract{Consorzio RFX and INFN-LNL have designed, built and operated the compact radiofrequency negative ion source NIO1 (Negative Ion Optimization phase 1) with the aim of studying the production and acceleration of H\textsuperscript{-} ions. In particular, NIO1 was designed to keep plasma generation and beam extraction continuously active for several hours. Since 2020 the production of negative ions at the plasma grid (the first grid of the acceleration system) has been enhanced by a Cs layer, deposited though active Cs evaporation in the source volume. For the negative ion sources applied to fusion neutral beam injectors, it is essential to keep the beam current and the fraction of co-extracted electrons stable for at least 1 h, against the consequences of Cs sputtering and redistribution operated by the plasma. The paper presents the latest results of the NIO1 source, in terms of caesiation process and beam performances during continuous (6÷7 h) plasma pulses. Due to the small dimensions of the NIO1 source ($20\  \mathrm{cm}$ x $\emptyset10\  \mathrm{cm}$), the Cs density in the volume is high ($10^{15} \div 10^{16} \  \mathrm{m}^{-3}$) and dominated by plasma-wall interaction. The maximum beam current density and minimum fraction of co-extracted electrons were respectively about $30$ A/m\textsuperscript{2} and $2$. Similarly to what done in other negative ion sources, the plasma grid temperature in NIO1 was raised for the first time, up to $80\ ^{\circ} \mathrm{C}$, although this led to a minimal improvement of the beam current and to an increase of the co-extracted electron current.}


\arxivnumber{2212.05801} 

\proceeding{8th International Symposium on Negative Ions, Beams and Sources\\
  2–7 October 2022\\
  Orto Botanico - Padova, Italy}

\begin{document}
\maketitle
\flushbottom

\section{Introduction}
\label{sec:intro}

\begin{figure}[htbp]
\centering 
\includegraphics[width=6 cm]{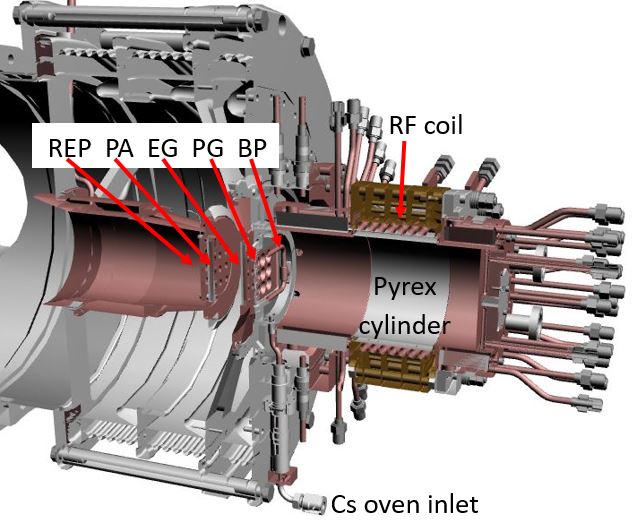}
\caption{\label{fig:nio} 3D section of the NIO1 source and acceleration system.}
\end{figure}

The NIO1 experiment, built and operated by Consorzio RFX and INFN-LNL, hosts a compact RF $H^-$ ion source, coupled to a 4-grid acceleration system. NIO1 aims at studying the source plasma and the beam production, to support the development of neutral beam injectors (NBI) for future fusion reactors (DEMO in particular).\cite{Cavenago2017, Cavenago2018, Cavenago2019, Cavenago2020} Differently from other larger facilities in this field, NIO1 was designed to withstand plasma generation and beam extraction for several hours.\cite{Serianni2020, Fantz2019, Wunderlich_2019} The structure of NIO1 is shown in the 3D section of fig. \ref{fig:nio}. The source volume is a $20\  \mathrm{cm}$ x $\emptyset10\  \mathrm{cm}$ cylinder; the plasma is sustained by $2-\mathrm{MHz}$ radiofrequency (maximum power $2.5\  \mathrm{kW}$), transmitted by a coil through a Pyrex cylinder. Apart from the Plasma Grid (PG), all the other surfaces are protected  by a Mo sheet. Besides the PG, the acceleration system is composed of an extraction grid (EG), post-acceleration grid (PA) and repeller grid (REP). Co-extracted electrons are mostly blocked on the EG thanks to magnets embedded in this grid. As in other tandem negative ion sources, the co-extraction of electrons and H\textsuperscript{-} stripping reactions are limited by a magnetic filter field, from $9\  \mathrm{mT}$ to $12.5\  \mathrm{mT}$ in proximity of the PG, as configuration F3 in ref. \cite{Cavenago2020}. The magnetic field is generated by an adjustable current flowing horizontally in the PG and by permanent magnets, some of which are externally placed and can be optionally removed. For a similar purpose, the PG and the bias plate (BP, an electrode in front of the PG) are positively and independently biased against the source body.

The negative ion sources for ITER and DEMO NBIs will be expected to reach an extracted ion current density in the order of $300\  \mathrm{A/m}^2$, and a fraction of co-extracted electrons below one.\cite{Serianni2020,Tran2022} This is only possible by evaporating Cs into the source; this lowers the works function of the PG surface, with an exponential increase of the negative ion production rate by surface reactions \cite{Bacal2015}. The work function depends on the purity and the thickness of the Cs layer deposited on the surface, which in turn depend not only on the evaporation rate but also on the Cs redistribution mechanisms caused by the hydrogen plasma and on the surface temperature \cite{Bacal2015,Friedl2017,Cristofaro_2020,Speth_2006,Ando1992,Yoshida2018,Sartori2018}. The management of Cs is then a critical task for any negative ion source aimed to continuous operation.  

In NIO1, Cs is evaporated by an oven, which mainly consists of a Cs reservoir, a manual protection valve and a duct leading to the source chamber; it is described in detail in refs. \cite{Barbisan2022,BarbisanNIO}.  While the overall Cs layer condition in the source can be indirectly monitored from the extracted beam current (the current flowing past the EG apertures) and from the co-extracted electron current (assumed equal to the EG current), the density of Cs circulating in the volume can be measured by a Laser Absorption Spectroscopy (LAS) diagnostic, with a line of sight parallel to the PG and at $19 \mathrm{mm}$ distance from it. \cite{BarbisanNIO} The purity and the conditions of the hydrogen plasma can be qualitatively monitored by Optical Emission Spectroscopy (OES), with a line of sight in axial direction from the rear side of the source. The effects on plasma luminosity ($200\div 1100\ \mathrm{nm}$ range) are also monitored by a photomultiplier, with a line of sight parallel to the OES one. The 2020 Cs experimental campaign in NIO1 allowed to increase the beam current and to reduce the co-extracted electrons \cite{BarbisanNIO}; however, the amount of Cs deposited into the source became too high, with a density of caesium $n_{cs}$ permanently stable at about $1\cdot10^{16} \  \mathrm{m}^{-3}$ during no-plasma phases. The consequences during plasma operation were: a reduction of beam current and an increase of the co-extracted electron current (as consequence of the lower H\textsuperscript{-} density at the PG and of plasma quasi-neutrality),  a degradation of voltage holding in the acceleration system, a reduction of the overall luminosity of the plasma in the visible-near infrared range (including the same Cs emission lines), a reduction of the $\mathrm{H}_\beta\mathrm{/H}_\gamma$ Balmer emission ratio.

The 2022 experimental campaigns, here presented, allowed to study the source and accelerator performances in continuous operation with more conservative Cs evaporation conditions compared to previous years. In the perspective of improving the beam current, it was proven in other negative ion sources that increasing the PG temperature improves the conditioning of the PG surfaces; temperature values from $80\div 100\ ^{\circ} $C up to $200\div 250\ ^{\circ}$C were reported as optimal.\cite{Sartori_2022,Speth_2006,Ando1992,Yoshida2018} In NIO1, given the present technical constraints of the cooling circuit and of vacuum components, it was possible to raise the PG temperature to about $85\ ^{\circ} $C, while keeping the source body at about $30\ ^{\circ} $C. Sec. \ref{sec:cs_evap} shows the results of the Cs conditioning of the NIO1 source in continuous plasma and beam extraction conditions, and the interplay between the magnetic filter field and the Cs-plasma conditioning. Sec. \ref{sec:PGheating} instead presents the effects of raising the PG temperature in NIO1.

\section{Cs evaporation}
\label{sec:cs_evap}
Due to the not easy accessibility of the Cs oven, its manual valve is opened at the beginning of each daily experimental session and closed only at its end. The evaporation rate from the Cs oven is mainly controlled by the reservoir temperature $T_{res}$, while valve and duct temperatures $T_{valve}$, $T_{duct}$ must be kept sufficiently higher (at least $T_{res}+40\div 50\ ^{\circ} $C) than the reservoir in order to avoid cold spots in the Cs drift path. During the 2022 experimental campaigns, $T_{res}$ was slowly increased in steps of $10\ ^{\circ}$C from the source body temperature, checking the detection of Cs from absorption and emission spectroscopy. As term of comparison, in the SPIDER experiment of the ITER neutral beam test facility \cite{Demuri2021,Sartori_2022}, $T_{res}$ is normally set in the three ovens between $105\ ^{\circ}$C and $155\ ^{\circ}$C to get Cs fluxes between $4$ mg/h and $20$ mg/h per oven, with Cs density values in the source volume (the expansion chamber is $1.8x0.25x0.8$ m large) up to few $10^{14} \  \mathrm{m}^{-3}$. In the more compact NIO1, no evidence of Cs from spectroscopy was detected for $T_{res}\leq 180\ ^{\circ}$C. Since the LAS detection threshold in NIO1 is about $1\cdot 10^{14} \  \mathrm{m}^{-3}$ this experimental evidence cannot be explained. Moreover no variation of electron and beam currents was detected. The cause of these phenomena are still under investigation. 

\begin{figure}[htbp]
\centering 
\includegraphics[width=15 cm]{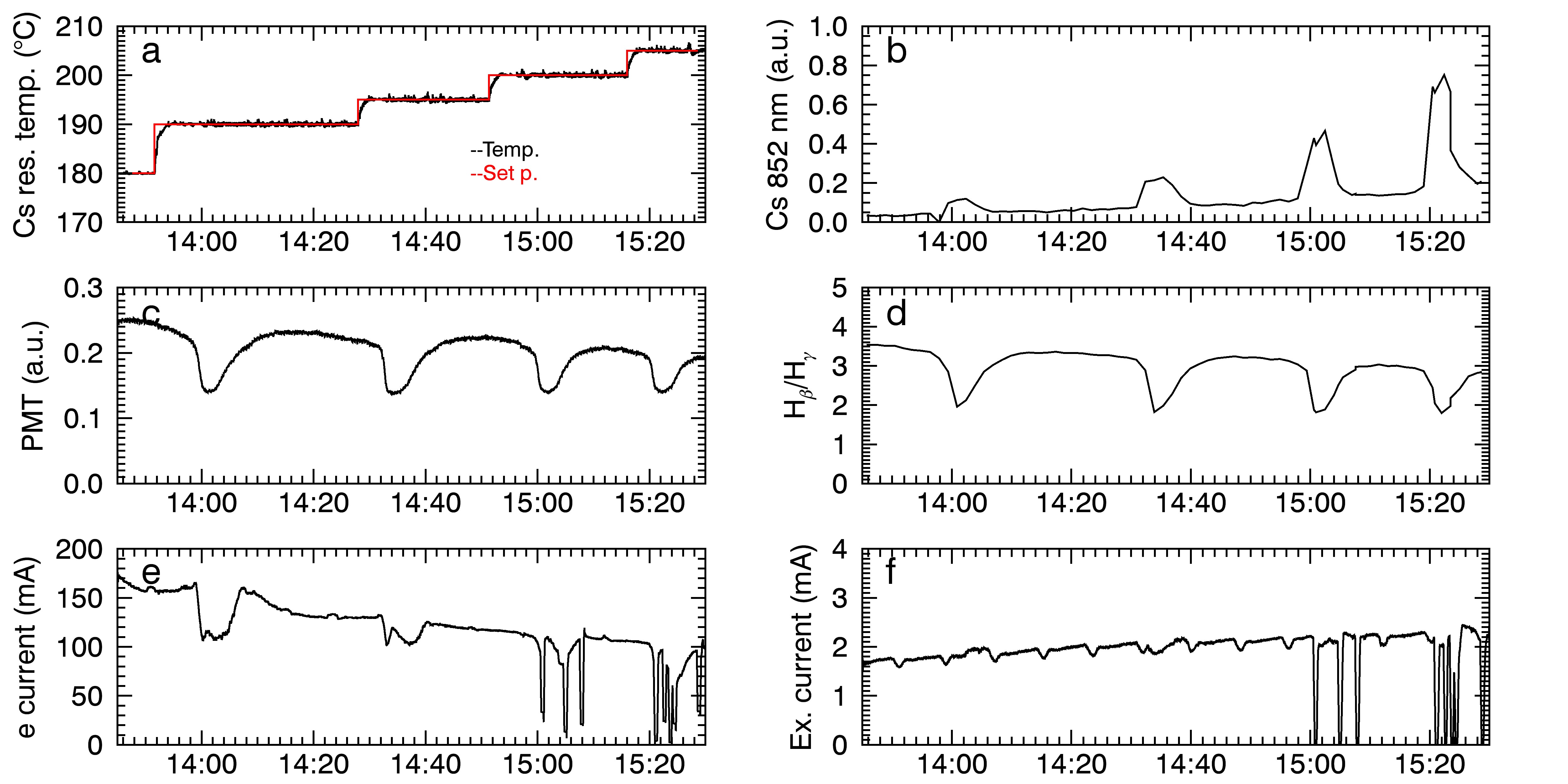}
\caption{\label{fig:firstcs} Evolution of the cesiation process, for the first time in 2022 in which Cs was detected. a) $T_{res}$ (measurements - black, set points - red), b) Emission intensity of the Cs $852$ nm line, c) plasma luminosity d) $\mathrm{H}_\beta \mathrm{/H}_\gamma$ emission intensity ratio, e) EG current and f) extracted beam current.}
\end{figure}

Fig. \ref{fig:firstcs} shows the effects on plasma and beam when Cs was detected for the first time by spectroscopic diagnostics. Fig. \ref{fig:firstcs}a shows the gradual increase of $T_{res}$ (measurements in black, set points in red), fig. \ref{fig:firstcs}b the emission intensity (in a.u.) of the Cs $852$ nm line. Fig. \ref{fig:firstcs}c shows the trend of the plasma luminosity as measured by the photo-multiplier, while fig. \ref{fig:firstcs}d shows the $\mathrm{H}_\beta\mathrm{/H}_\gamma$ Balmer emission intensity ratio measured by OES. Lastly, fig. \ref{fig:firstcs}e and \ref{fig:firstcs}f show the co-extracted electron current and the extracted beam current, respectively; the spikes in the signal refer to breakdowns in the accelerator, rapidly recovered. NIO1 was operated in the following conditions: RF power $P_{RF}=1.4$ kW, source pressure $p_{source}=0.75$ Pa, PG-source voltage bias $V_{PGS}=30$ V, BP-source body bias current $I_{BPS}=0.3$ A, PG filter current $I_{PG}=10$ A, extraction voltage (EG-PG) $U_{ex}=0.6$ kV, total acceleration voltage (PA-PG) $U_{tot}=6$ kV. 

Each time $T_{res}$ is increased, the Cs evaporation rate (and then the Cs density) into the source changes with a delay of about $5\div 7$ min., resulting not only in a sudden increase of the $852$ nm Cs line, but also in a reduction of plasma luminosity and of the $\mathrm{H}_\beta\mathrm{/H}_\gamma$ ratio, as witnessed by the same NIO1 in 2020 and in other machines in case of overcesiation.\cite{BarbisanNIO} The reduction of $\mathrm{H}_\beta \mathrm{/H}_\gamma$ can be interpreted as a reduction of electron density or an increase of electron temperature.  These phenomena, however, are recovered in few minutes, as if the balance between Cs influx, source volume and surfaces were able to restore an equilibrium, removing most but not all the Cs from the volume. With the progress of Cs evaporation, the $852-nm$ Cs line grows, while the plasma luminosity and the $\mathrm{H}_\beta \mathrm{/H}_\gamma$ ratio decrease. Correspondingly and as expected, the electron current decreases and the beam current increases. The electron current follows the relatively rapid evolution of Cs evaporation, while the ion current is not affected by the Cs density spikes.

During each day of Cs evaporation, after hours of active evaporation $T_{res}$ was set below $180\ ^{\circ}$ C during the plasma phase, to check the consequent reduction of the Cs density. It was possible to steadily operate the oven up to $T_{res}=215\ ^{\circ}$ C, with the possibility to decrease the Cs density back during the day; every day following a plasma session, the Cs density before the initiation of the plasma discharge  was below the detection threshold of LAS. As comparison, in the overcesiation case of the 2020 experimental campaign, $T_{res}$ was set between $220\ ^{\circ}C$ and $240\ ^{\circ}$ C, with a permanent Cs density in the order of $10^{16} \  \mathrm{m}^{-3}$. 

\begin{figure}[htbp]
\centering 
\includegraphics[width=15 cm]{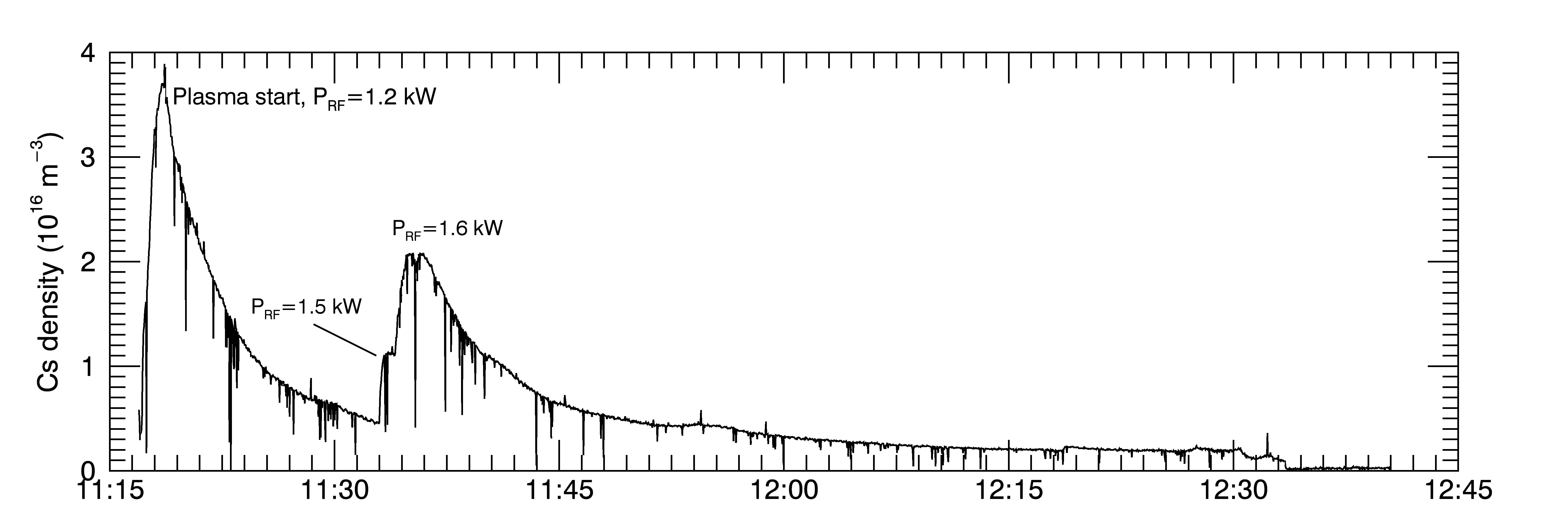}
\caption{\label{fig:sputtering} Cs density, measured by LAS, as a function of time, in a short plasma session. The Cs oven was not evaporating ($T_{res}<165\ ^{\circ}$ C).}
\end{figure}

During the continuous plasma operation, which normally lasted up to $7\div 8\ h$, the Cs density measured by LAS typically ranged between $1\cdot 10^{15} \  \mathrm{m}^{-3}$ and $1\cdot 10^{16} \  \mathrm{m}^{-3}$ with active evaporation. The Cs density in the volume is not just a function of $T_{res}$, but it's strongly influenced by the plasma, which ionizes cesium (Cs\textsuperscript{+} and the excited states of Cs are not detectable by LAS \cite{Fantz2011}), but also removes Cs from the surfaces. Fig. \ref{fig:sputtering} shows the plasma effects, in a short experimental session, with the Cs oven not effectively evaporating ($T_{res}<165\ ^{\circ}$ C). The Cs density measured by LAS is plotted against time; spikes are just an artifact of data analysis. The operative conditions were: $P_{RF}=1.2\mathrm{,}\ 1.5\mathrm{,}\ 1.6$ kW (see plot labels), $p_{source}=0.75$ Pa, $V_{PGS}=7,\ 19$ V and $I_{BPS}=0.3\mathrm{,}\ 0$ A (plasma potential above power supply setting) at low and high $P_{RF}$, respectively, $I_{PG}=400$ A (more filtering to contrast the increase of electron current with decesiation). As shown in Fig. \ref{fig:sputtering}, the sputtering alone can temporarily raise the Cs density up to some  $1\cdot 10^{16} \  \mathrm{m}^{-3}$, then decreasing in $15\div 30$ min.; the effect is dependent on $P_{RF}$. Such high values of Cs density can be explained by the fact that NIO1 is much smaller than other negative ion sources for fusion, then the surface-to-volume ratio is much higher and so the effect of sputtering. The data of fig. \ref{fig:sputtering} indicate also that, in this range of source dimensions, hypothetical plasma phases of few minutes or few seconds would have completely different Cs dynamics than continuous operation, which is the target for possible future DEMO neutral beam injectors.

\begin{figure}[htbp]
\centering 
\includegraphics[width=15 cm]{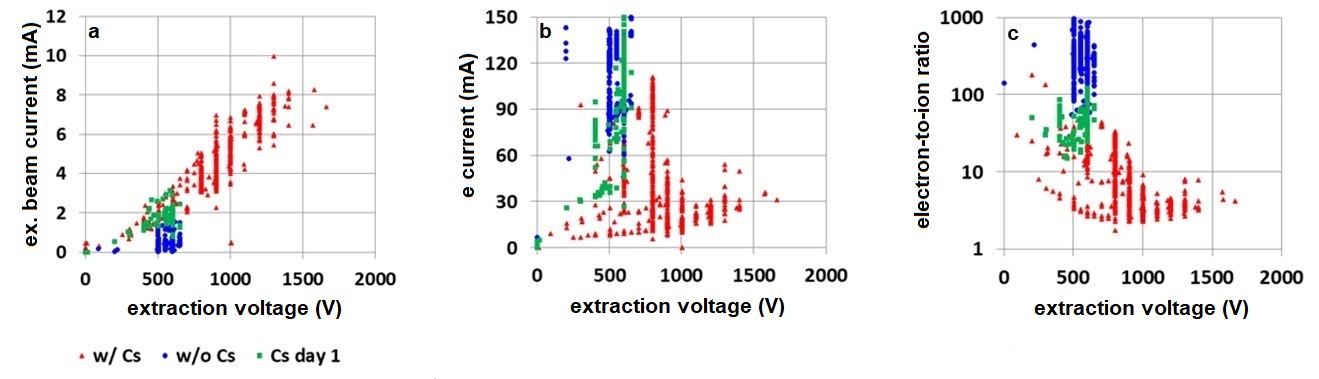}
\caption{\label{fig:beam_evap} a) Extracted beam current b) EG current and c) fraction of co-extracted electrons, as a function of the extraction voltage. The measurements were taken before (blue dots) and during (red dots) the evaporation of Cesium. Green dots refer to the very first day of Cs evaporation.}
\end{figure}

The improvements due to cesiation on beam properties are shown in fig. \ref{fig:beam_evap}. The extracted beam current (a), the co-extracted electron current (b) and the fraction of co-extracted electrons are plotted against the extraction voltage; every case is a dataset, i.e. a point in time in which all the diagnostic and plant information was saved as a relevant condition during the experimental sessions. Blue points refer to pre-cesiation activities, green dots to the first day of cesiation and red points to the following days. The evaporation of Cs allowed a higher production of negative ions and then a higher beam current. Because of space charge, part of the electrons in proximity of the PG were repelled, with a consequent lower co-extracted electron current. This had the important benefit of improving the high voltage stability of the acceleration system, allowing to increase $U_{ex}$ and then extract the larger density of negative ions available. Increasing $U_{ex}$ with insufficient availability of H\textsuperscript{-} and then beam current would also have led to improper beam optics, as well known by perveance physics. \cite{Forrester1988} Comprehensively, the Cs evaporation allowed to reach a maximum extracted beam current density of about $25\ \mathrm{A/m}^2$ ($1\ \mathrm{mA}$ of beam current corresponds to about $2.7\ \mathrm{A/m}^2$) and a fraction of co-extracted electrons down to about $2$. 

\begin{figure}[htbp]
\centering 
\includegraphics[width=15 cm]{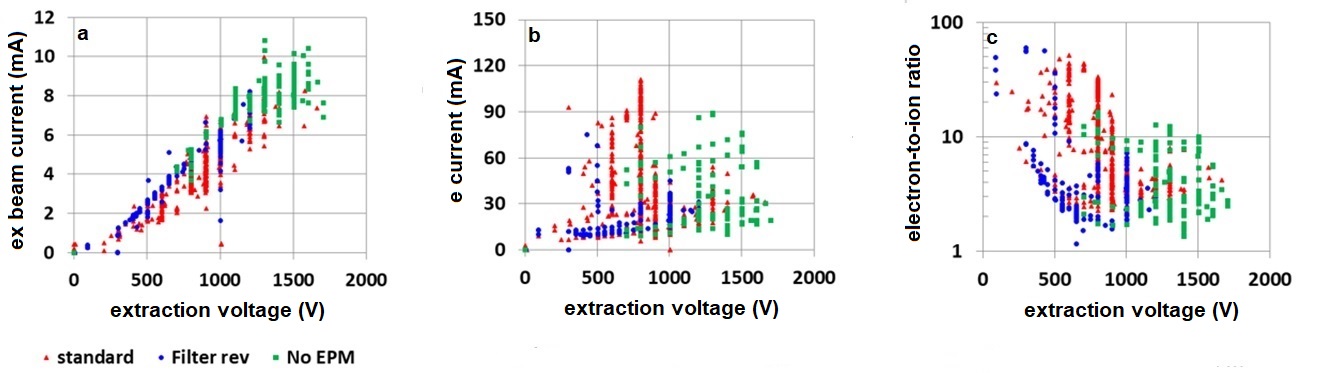}
\caption{\label{fig:beam_mag} a) Extracted beam current  b) EG current and c) fraction of co-extracted electrons, as a function of the extraction voltage.$I_{PG}$ was in standard configuration (red, reinforcing the filter field) or in reversed one (blue, weakening the field). The cases in green dots have $I_{PG}$ as in standard configuration, external permanent magnets are absent.  }
\end{figure}

To further improve the beam current, attempts were made to improve the plasma density in front of the PG, and then the production rate of negative ions, by weakening the magnetic filter field. In the standard configuration, the magnetic filter field is directed downwards and is generated by permanent magnets in the modified multipole system (peak $5\ \mathrm{mT}$), plus external permanent magnets (peak $4\ \mathrm{mT}$), plus $I_{PG}$ (up to $3.5\ \mathrm{mT}$).\cite{Cavenago2020} Fig. \ref{fig:beam_mag} shows the effect of weakening the filter field on the extracted beam current, on the EG current and on the fraction of co-extracted electrons, for several values of $U_{ex}$. Red dots refer to datasets acquired in the standard configuration (as in fig. \ref{fig:beam_evap}), while with the blue dots the $I_{PG}$ direction (and its magnetic field) were reversed, weakening the field. Lastly, in the cases in green $I_{PG}$ was in standard direction, but no external permanent magnets were present. The results of fig. \ref{fig:beam_mag} show that weakening the magnetic filter field has a slighty beneficial effect on the extracted beam current, which can reach $30\ \mathrm{A/m}^2$, with negligible improvements on the fraction of co-extracted electrons. For reasons to be investigated, the absence of the external magnets allowed to safely increase $U_{ex}$. The magnetic filter field intensity however had a more complex influence of the cesiation process: while the standard configuration allows to condition the source to a stable level within the timescale of the hour at $p_{source}=0.75$ Pa, without the external magnets the source Cs conditioning was much slower, and in any case it required to increase $p_{source}$ up to $1.25\div 1.5$ Pa to make it happen. This may be an effect of the rise of the electron temperature with lower magnetic field, which leads to a higher sputtering on the surfaces. Increasing the source pressure in the Pa order has the known effect of reducing the electron current, compensating the lower magnetic field intensity. \cite{Tonks1929}
\section{Effects of PG heating}
\label{sec:PGheating}

\begin{figure}[htbp]
\centering 
\includegraphics[width=10 cm]{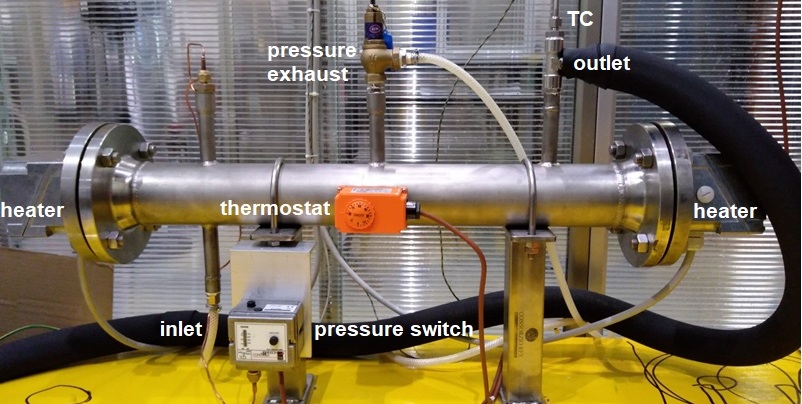}
\caption{\label{fig:heater} Components of the water heater for the control of the PG temperature.}
\end{figure}

To improve the beam performances described in fig. \ref{fig:beam_evap}, as mentioned in sec. \ref{sec:intro}, the water cooling system was modified in order to increase the PG temperature $T_{PG}$; the entire source can be normally put at no more than $30\ ^{\circ}$ C (the water cooling also serves the RF generator, which has safety limits on water flux and temperature). To heat the PG, a water heater was designed and built; it is shown in fig. \ref{fig:heater}. It's a stainless steel cylinder of $8$ cm inner diameter and $80$ cm length, with a $4$-kW set of three resistors (for 3-phase $380$-V power supply) on each opposite end. Each resistor can be selectively activated/deactivated, and is modulated by means of solid state relays. These are controlled by a PID system, which takes the temperature feedback by a thermocouple, axially immersed in the pipe which connects to the outlet hose. The safety is guaranteed by a thermostat and a pressure switch, which can stop all the resistors, plus a pressure exhaust. The heater inlet is connected to the main cold-water line of the cooling system. The heater outlet is thermally insulated in foam. The hose reaches the high voltage deck (HVD) of the ion source as an insulated coil, to contrast the still minimal water conductivity. Once flown trough the PG, the hot water flows in series through the BP and is finally mixed in the HVD water return collector. The flux on the heater line is typically $1.6$ L/min. of water at a temperature up to nearly $100\ ^{\circ}$C. The decrease of  temperature along the hot water hose grows with temperature itself, and can reach about $15\ ^{\circ}$C.

\begin{figure}[htbp]
\centering 
\includegraphics[width=15 cm]{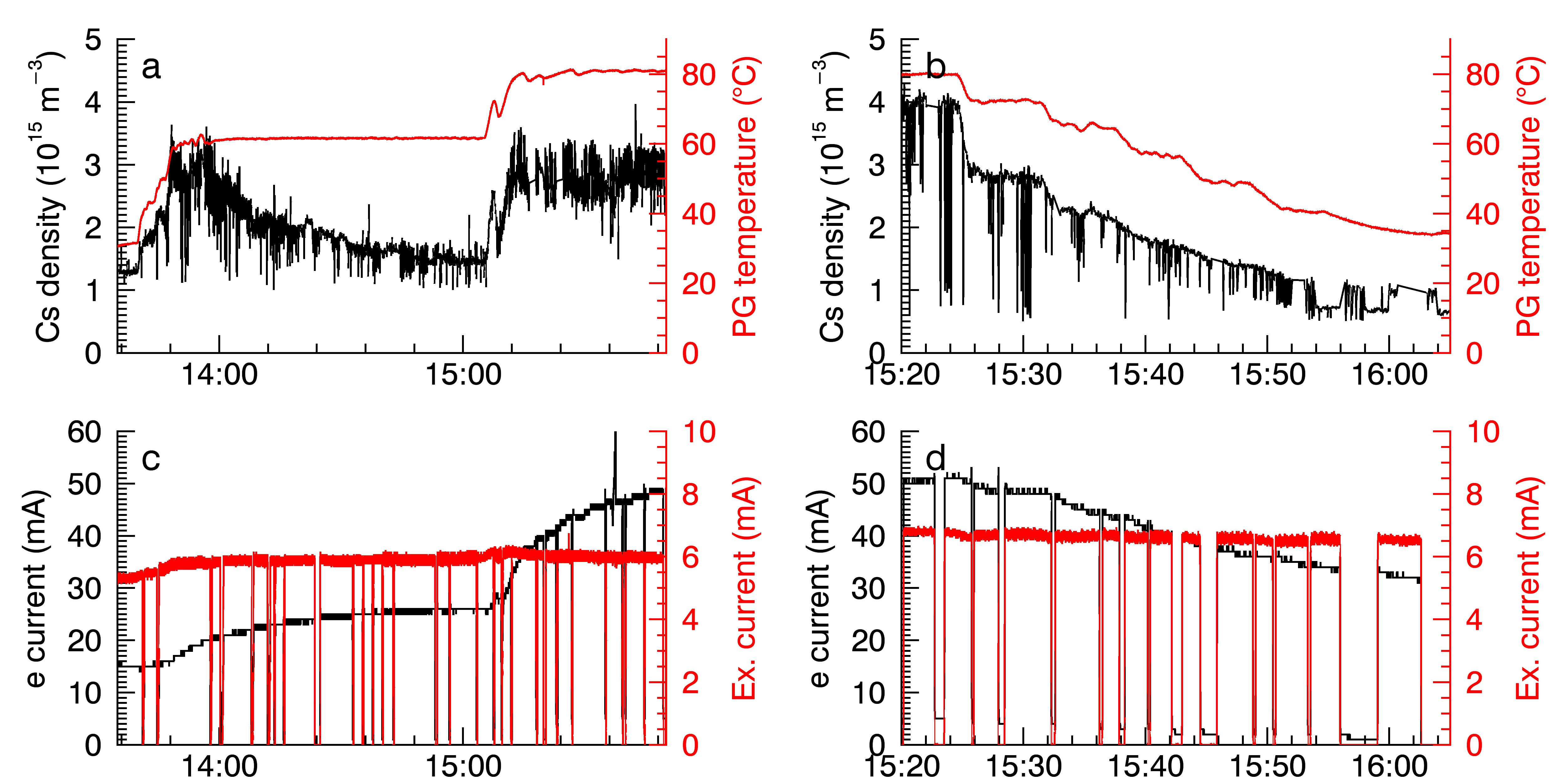}
\caption{\label{fig:PGH} Effects of increasing (plots a and c) and decreasing (plots b and d) $T_{PG}$. a,b) Cs density measured by LAS (black) and $T_{PG}$ (red) for the two experimental cases, as a function of time. c,d) co-extracted electron current (black) and the extracted beam current (red) as a function of time.}
\end{figure}

Fig. \ref{fig:PGH} shows the effects of changing the PG temperature on two experimental cases, one with increasing $T_{PG}$ (plots a and c) and one with decreasing $T_{PG}$ (plots b and d). Plots a and b show the Cs density measured by LAS (black) and the PG temperature (red) as measured at the PG water outlet, while plots c and d show the co-extracted electron current (black) and the extracted beam current (red), as a function of time; the large jumps in the currents are due to breakdown in the PG-EG gap or in the EG-PA gap. The operative conditions for the first case (fig. \ref{fig:PGH}a,c) were: $P_{RF}=1.6$ kW, $p_{source}=0.75$ Pa, $V_{PGS}=45$ V, $I_{BPS}=0.15$ A, PG filter current $I_{PG}=400$ A, $U_{ex}=1.3$ kV, $U_{tot}=13$ kV; $T_{res}$ was at $195\ ^{\circ}$C until 13:59, and then lowered to $180\ ^{\circ}$C (negligible evaporation). The operative conditions for the second case (fig. \ref{fig:PGH}b,d) were similar, except $U_{ex}=1.5$ kV, $U_{tot}=15$ kV and $T_{res}<180\ ^{\circ}$C in the whole time interval. In both cases $I_{PG}$ was in standard configuration, with the external permanent magnets.
The first experimental case shows that heating the PG up to $80\ ^{\circ}$C can lead to a significant desorption of Cs into the volume, in the same extent of the Cs oven: stopping the evaporation after 13:59 causes a reduction of the Cs density which is slightly lower than the increase in Cs density due to the PG heating. However, increasing $T_{PG}$ does not seem to improve the beam properties: the co-extracted electron current even increases with $T_{PG}$, while the beam current undergoes minimal increases; this effect may be also mixed with a slight increase of the dark current between source-HVD and ground, due to the increase of  water conductivity with temperature. The second case is coherent with the first one: decreasing $T_{PG}$ leads to a reduction of the Cs density in the volume and of the electron current, with minimal variations of the beam current.  

\section{Conclusions}
In the experimental campaigns of 2022, Cs was evaporated in the NIO1 source in a more controlled way. While the phenomena observed in spectroscopic data are compatible with the overcesiation conditions encountered in 2020 \cite{BarbisanNIO}, the cesiation process was fully reversible. The LAS data showed that the Cs density evolves in a timescale of about half hour, and apart from the Cs oven input it is dominated by plasma-surface interactions. This phenomenon, together with the relatively high values of Cs density, is to be probably attributed to the small dimensions of the source and then to the high surface-to-volume ratio. The magnetic filter field intensity can affect the timing of the cesiation process and the lowest pressure at which the source can be operated. Contrarily to the expectations, raising the PG temperature in the $30\div 80\ ^{\circ}$C range in the NIO1 source conditions gave null to negative beam improvements. Future works in NIO1 will aim to understand and solve the causes of the relatively low values of beam current density (one order of magnitude below the ultimate target) and to reduce the fraction of co-extracted electrons below $1$.


\acknowledgments
This work has been carried out within the framework of the EUROfusion Consortium, funded by the European Union via the Euratom Research and Training Programme (Grant Agreement No 101052200 — EUROfusion). Views and opinions expressed are however those of the author(s) only and do not necessarily reflect those of the European Union or the European Commission. Neither the European Union nor the European Commission can be held responsible for them. This work has also been carried out within the framework of the ITER-RFX Neutral Beam Testing Facility (NBTF) Agreement and has received funding from the ITER Organization. The views and opinions expressed herein do not necessarily reflect those of the ITER Organization. This work was also supported in part by the Swiss National Science Foundation, by INFN-CSN5 experiment Ion2neutral an by project INFN-E.


\bibliographystyle{JHEP}
\bibliography{refs}







\end{document}